\documentclass[twocolumn]{aa}
\usepackage{graphicx}
\usepackage{txfonts}
\def\brfrac#1#2{\left(\dfrac{#1}{#2}\right)}
\def\dfrac#1#2{{\displaystyle\frac{\mathstrut #1}{#2}}}

\def\ea{{\rm et~al.\ }}
\def\rmAA{{\rm \AA}}
\def\Mdot{{\dot{M}}}

\def\Mbh{{M_{\rm BH}}}
\def\Msun{{M_\odot}}
\def\Lnu{{L_\nu}}

\def\Ledd{{L_{\rm Edd}}}
\def\Leddc2{{L_{\rm Edd}/c^2}}
\def\Teff{{T_{\rm eff}}}

\def\rs{{r_{\rm Sch}}}
\def\rsg{{r_{\rm sg}}}
\def\rtrap{{r_{\rm trap}}}
\def\rsga{{r_{\rm sg, a}}}
\def\rsgb{{r_{\rm sg, b}}}
\def\rsgc{{r_{\rm sg, c}}}

\def\lsg{{\lambda_{\rm sg}}}

\begin{document}
   \title{The origin of optical emission from super-Eddington
       accreting Active Galactic Nuclei: the case of \object{Ton~S~180}}


   \author{Toshihiro Kawaguchi\inst{1,}\inst{2}, 
          Arnaud Pierens\inst{1}
          \and
          Jean-Marc Hur\'e\inst{1,}\inst{3}
          }

   \offprints{T. Kawaguchi}
   \authorrunning{T. Kawaguchi et al.}
   \titlerunning{Optical emission from super-Eddington accreting AGN}

   \institute{LUTh/Observatoire de Paris-Meudon et CNRS UMR 8102, 
              Place Jules Janssen, 92195 Meudon Cedex, France \\
             \email{Toshihiro.Kawaguchi@obspm.fr; Arnaud.Pierens@obspm.fr; 
                    Jean-Marc.Hure@obspm.fr}
             \and
             Postdoctoral Fellow of the Japan Society for the 
              Promotion of Science
             \and
             Universit\'e Paris 7 Denis Diderot, 2 Place Jussieu, 
              F-75251 Paris Cedex 05, France
             }

   \date{Received 2 May 2003 / Accepted 17 October 2003}

   \abstract{
{ 
Self-gravitating accretion discs have only been studied in a 
few nearby objects using maser spots at the parsec-scale.
We find a new spectral window for observing the self-gravitating 
accretion disc in super-Eddington accreting Active Galactic Nuclei
(AGNs).
This window is determined by calculating the outermost radius 
($\rsg$) of a non self-gravitating disc and the corresponding 
emission wavelength ($\lsg$) as a function of various disc parameters.
We find that 
$\lsg$ reaches 
$\sim 4000 \rmAA$ for $\alpha=0.1$,
when $\Mdot \gtrsim 70 \, (\Mbh / 10^7 \Msun)^{-1} \, \Leddc2$
(where $\alpha$, $\Mdot$, $\Mbh$ and $\Ledd$ are, respectively, 
the viscosity parameter, gas accretion rate onto the central black hole (BH),
the BH mass and the Eddington luminosity).
Moreover, $\lsg$ is as small as $\sim 1500 \rmAA$ for
$\alpha =0.001$, which is 
the smallest $\alpha$ case in this study.
Therefore, 
the window for observing the self-gravitating part of an 
AGN accretion disc is from $\sim 2 \mu$m to $\lsg$.
Incidentally, 
$\rsg$ can be less than the photon trapping radius 
for $\Mdot \gtrsim 10^{3.3} \Leddc2$. 
Namely, a self-gravitating, 
optically-thick, advection-dominated accretion disc
is expected to appear in the extremely high accretion rate regime.

Next, we demonstrate that the Mid-Infrared to X-ray spectrum 
of a bright, well-studied Narrow-Line Seyfert 1 galaxy, Ton~S~180, 
is indeed well fitted by the spectrum arising from 
the following three components: an inner slim disc (with a corona), 
an outer, self-gravitating non-Keplerian disc 
and a dusty torus. 
The total mass, BH mass plus the entire disc mass, 
is found to be about $(1.4 - 8.0) \Mbh$.
If the surface density varies with radius $r$ in proportion to 
$r^{-0.6}$, the total mass 
is consistent
with the central mass estimated by H$\beta$ and [O III] widths.
}
   }

   \maketitle
%

\section{Introduction}

The Optical/UV/X-ray emission from 
Active Galactic Nuclei (AGNs) is thought to arise from a hot 
accretion disc around a super-massive Black Hole 
(BH; e.g., Mushotzky, Done \& Pounds 1993; 
{ Koratkar} \& Blaes 1999), 
while the { Mid- and Near Infrared (IR) radiation comes 
from a cold}, dusty torus (e.g., Telesco \ea 1984; Thatte \ea 1997; 
Pier \& Krolik 1993). 
Disc self-gravity can generally be ignored in the innermost region
where the high energy spectrum is formed. At large distances 
(typically $\gtrsim 10^3 - 10^4$ Schwarzschild radii, depending on 
accretion parameters), the disc mass is expected to play a significant 
role. 
Disc self-gravity has been tested only at the parsec-scale via 
maser spots detected in a few nearby objects 
(e.g. Nakai, Inoue \& Miyoshi 1993; Miyoshi \ea 1995; 
Hur\'e 2002; Lodato \& Bertin 2003).

Narrow Line Seyfert 1 galaxies (NLS1s) and their high luminosity 
analogue, 
Narrow-Line QSOs, are supposed to have high accretion rates 
among the AGN population (e.g, Brandt \& Boller 1998; 
{ Mineshige \ea 2000}). 
The gas accretion rate $\Mdot$ for { those} objects 
{ is} expected to
be comparable to or larger than $10 \Leddc2$,
where $\Ledd$ is the Eddington luminosity.
For such super-Eddington accretion rates, 
the outer edge of the non self-gravitating part of the accretion disc
{ radiates} optical { continuum} 
emission ($\sim 4000 \rmAA$) for a certain 
parameter set (Kawaguchi 2003).
Thus, they are potentially good candidates to study the 
self-gravitating part of the disc observationally at longer wavelengths 
(Collin et al. 2002; Kawaguchi 2003). 

In this paper, we qualitatively describe 
the wavelength corresponding to the emission from the outer edge of 
a non self-gravitating disc for various accretion parameter sets.
{ This is done in \S 2.}
We present { in \S 3 }
a Mid-IR to X-ray spectral modeling of a bright, well-studied 
NLS1, Ton~S~180 (PHL912), 
for which numerous, multi-waveband observations 
are available (e.g., Wisotzki \ea 1995; Turner et al. 2002; 
Vaughan et al. 2002). 
In \S 4, several discussions are presented.
The final section is devoted to a summary.


\section{ Critical Radius and Critical Wavelength}

In this section, we calculate the radius of the 
outer edge of the non self-gravitating { slim} disc ($\rsg$), i.e.
inner edge of the self-gravitating disc,
as a function of  the BH mass $\Mbh$, $\Mdot$ and 
viscosity parameter $\alpha$. We will then show the 
corresponding wavelength ($\lsg$) that is relevant to emission from $\rsg$.
We define $M_7$ for $\Mbh$ in the unit of $10^7 \Msun$.

\begin{figure}
   \centering
    \includegraphics[width=8.5cm]{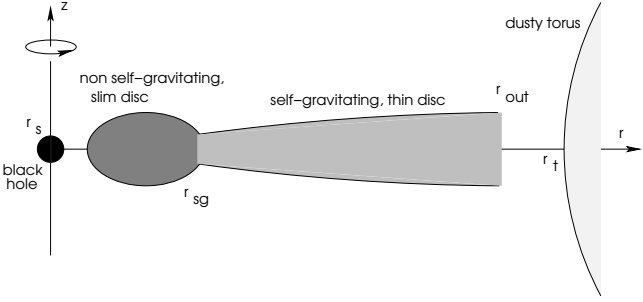}
    \caption{Schematic configuration (not to scale) of the three 
    relevant components surrounding the black hole: a slim disc, 
    a self-gravitating disc and a torus.
    For Ton~S~180 (\S 3), the characteristic radii are found to be the 
    followings: 
    $\rsg \simeq 3 \times 10^3 \rs$, $r_{\rm out} \simeq 3 
    \times 10^4 \rs$ and $r_{\rm t} \simeq 3 \times 10^5 \rs$.
    The broad-line region is located at $r \simeq 1.4 
    \times 10^5 \rs$ (\S 4).
}
         \label{fig:config}
   \end{figure}

Figure~\ref{fig:config} illustrates the configuration considered
in this study.

%
%

\subsection{Critical Radius}

For super-Eddington accretion rates (i.e. $\Mdot \gtrsim 10 \Leddc2$), an optically-thick, advection-dominated regime appears (Abramowicz \ea 1988). Inside a certain radius $\rtrap \sim 0.5 \Mdot (\Leddc2)^{-1} \rs$ (Begelman \& Meier 1982; Kawaguchi 2003), 
the timescale for photon diffusion is larger than the accretion timescale,
and thus most of the photons emitted inside the flow are trapped within it.
Thus, { advection cooling becomes} dominant over radiative cooling
at $r < \rtrap$. 

We define $\rsg$ as the radius where the mass density in the disc mid-plane 
($\rho_{\rm mid-plane}$) equals $\rho_{\rm sg}$.
Here, $\rho_{\rm sg}$ is a critical density above 
which 
self-gravity of the disc 
must be considered:
\begin{equation}
\label{eq:rho_sg}
\rho_{\rm sg} =\Omega_{\rm K}^2 / (4 \pi {\rm G}) = 
2.2 \, M_7^{-2} (r / 3 \rs)^{-3} {\rm g/cm}^3,
\end{equation}
where $\Omega_{\rm K}$ is the Keplerian rotation frequency 
(e.g., Goldreich \& Lynden-Bell 1965; Hur\'{e} 1998). 
Inside $\rsg$, $\rho_{\rm mid-plane}$ is smaller than $\rho_{\rm sg}$,
thus the disc self-gravity is negligible there.

{ Many authors have tried to derive $\rsg$ 
(e.g., Shore \& White 1982).
Actually, $\rsg (\Mbh, \, \Mdot, \, \alpha)$ derived and used in this 
study is qualitatively consistent with that in Hur\'{e} (1998).}

In order to derive 
$\rsg$, we use analytical formulae for the mass density 
($\rho_{\rm mid-plane}$) given for a Newtonian gravitational potential 
(e.g., Kato, Fukue \& Mineshige 1998). 
We confirm that these standard-disc formulae for various physical quantities 
are consistent with numerical computations (Kawaguchi 2003) within a factor 
of $\sim 1.5$ at any radius larger than $\rtrap$.
In other words, the flow behaves like a standard disc 
(Shakura \& Sunyaev 1973) outside $\rtrap$,
even { for} a super-Eddington accretion rate.

\begin{figure}
   \centering
    \includegraphics[width=8.5cm]{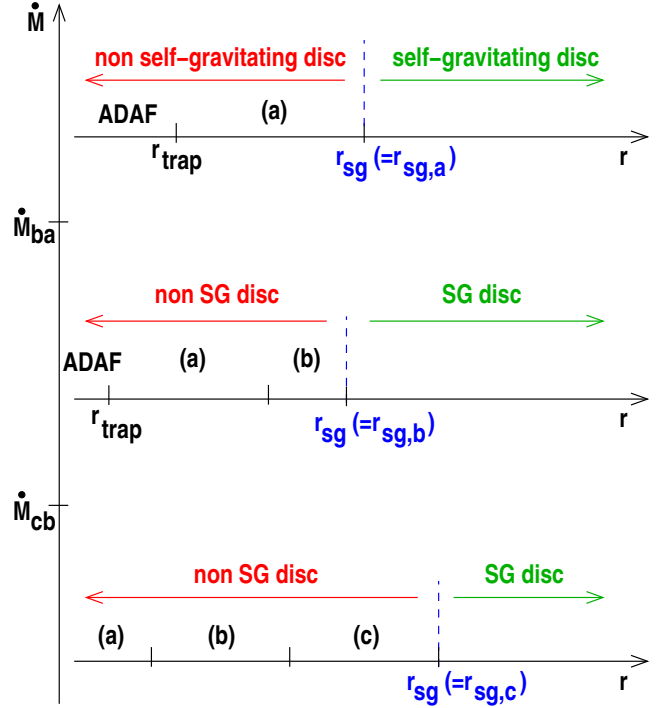}
    \caption{ 
    Classification of the region where $\rsg$ appears (\S 2.1).
    Region (c) harbors $\rsg$ when $\Mdot$ is less than $\Mdot_{\rm cb}$,
    while regions (b) and (a) are, respectively, in charge of $\rsg$
    when $\Mdot$ is smaller and larger than $\Mdot_{\rm ba}$.
    Around the radius $\rtrap$, the flow undergoes a change from
    radiation cooling-dominated regime to advection-dominated
    regime (optically-thick ADAF).
    }
         \label{fig:rsg}
   \end{figure}

{ In a standard disc, the following}
three regions must be distinguished { according to 
the sources of the pressure and opacity} 
(e.g., Shakura \& Sunyaev 1973; Kato \ea 1998).
{ In which region $\rsg$ appears is determined by $\Mdot$ 
(see Figure~\ref{fig:rsg}).}
%
%
\begin{itemize}
\item  
an outer region (c), where the gas pressure and absorption opacity 
dominate over radiation pressure and electron-scattering opacity, 
respectively. 
With a relatively low accretion rate, 
i.e. $\Mdot \lesssim \Mdot_{\rm cb} = 2 \, M_7^{-1}
(\alpha / 0.1)^{7/13} \, \Leddc2$,
$\rsg$ appears at the outer region (c):
\begin{equation} 
\label{eq:rsg_c}
 \rsgc 
 = 6700 \, \brfrac{\alpha}{0.1}^{28/45} M_7^{-52/45} 
         \brfrac{\Mdot}{\Leddc2}^{-22/45} \rs.
\end{equation}
\item 
a middle region (b) where gas pressure and electron-scattering opacity
are dominant. 
For a moderately high accretion rate, i.e. 
$\Mdot \lesssim \Mdot_{\rm ba} = 70 \, M_7^{-1} 
(\alpha / 0.1)^{0.4} \, \Leddc2$,
$\rsg$ is given by: 
\begin{equation}
\label{eq:rsg_b}
 \rsgb 
 = 5700 \, \brfrac{\alpha}{0.1}^{14/27} M_7^{-26/27} 
         \brfrac{\Mdot}{\Leddc2}^{-8/27} \rs,
\end{equation}
\item 
an inner region (a) 
which is described by radiation pressure and electron-scattering opacity. 
For a higher accretion rate ($\Mdot \gtrsim \Mdot_{\rm ba}$), 
$\rsg$ is described as 
\begin{eqnarray}
\label{eq:rsg_a}
 \rsga  
 = 250 \, \brfrac{\alpha}{0.1}^{2/9} M_7^{-2/9} 
         \brfrac{\Mdot}{\Leddc2}^{4/9} \rs.
\end{eqnarray}

\end{itemize}

\begin{figure*}
   \centering
    \includegraphics[angle=-90,width=\textwidth]{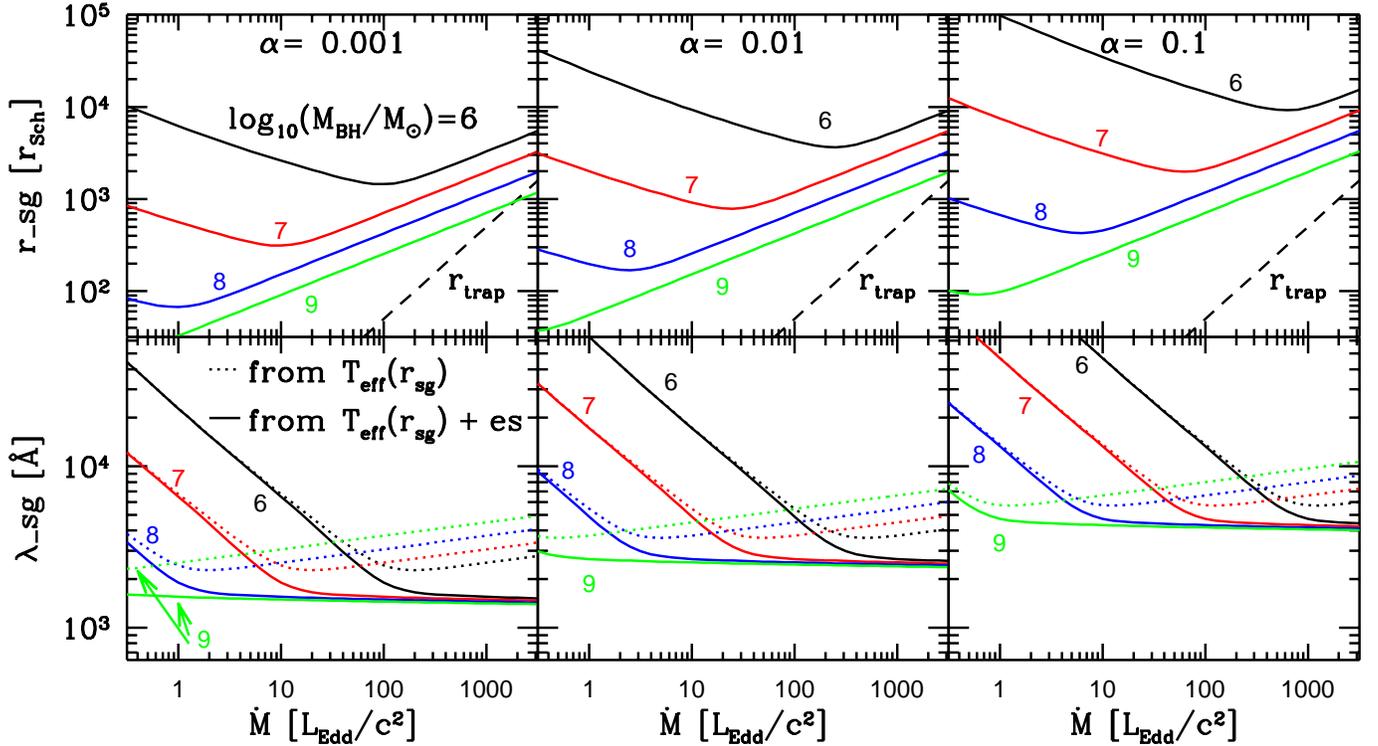}
   \caption{Critical radius and wavelength 
    [$\rsg$ (upper) and $\lsg$ (lower)] for various $\Mdot$ and $\alpha$.
    Here, $r_{\rm trap}$ (dashed lines in the upper panels) is the 
    radius inside which photon trapping plays a role.
    The dotted lines in the lower panel are computed based on 
    the assumption that the radiation from $\rsg$ is blackbody,
    while solid lines take the effect of electron scattering (modified
    blackbody) into account.
    If $\lsg \lesssim 2 \mu$m, it means that we are able to observe 
    the self-gravitating accretion disc as Near-IR/optical continuum
    radiation.
           }
              \label{fig:rlambda3}
    \end{figure*}

Given $\Mbh$ and $\Mdot$, the critical radius $\rsg$ can be 
written as Max($\rsgc, \rsgb, \rsga$), 
for the whole range of accretion rates.
Instead of a piece-wise function, we employ a simple formula: 
$\rsg = (\rsga^3 + \rsgb^3 + \rsgc^3)^{1/3}$.
Figure~\ref{fig:rlambda3} (upper panels) shows $\rsg$ for 
various $\Mbh$, $\Mdot$ and $\alpha$. 
Until the accretion rate reaches $\Mdot_{\rm ba}$,
$\rsg$ decreases with an increasing $\Mdot$,
while $\rsg$ changes the sign of its $\Mdot$-dependency at 
$\Mdot \sim \Mdot_{\rm ba}$. 

We find that with { $\Mdot \gtrsim 10^{3.3} \ (\alpha / 0.001)^{0.4}$ 
$(\Mbh / 10^9 \Msun)^{-0.4} \ \Leddc2$}, 
$\rsg$ can be less than $r_{\rm trap}$.
The above expressions are no longer valid in 
the regime in which $\rsg < \rtrap$.
In that case, 
a self-gravitating, 
optically-thick, advection-dominated accretion disc appears.
{ Therefore, we restrict ourselves to}
accretion rates lower than $10^{3.5} \Leddc2$,
{ in order to ensure that $\rsg \gtrsim \rtrap$}.

%
%

\subsection{Critical Wavelength}

Next, we evaluate the effective temperature at $\rsg$, 
and subsequently a critical wavelength $\lsg$ corresponding to the 
emission from $\rsg$. 
At $r > \rtrap$, the effective temperature $\Teff$ at radius $r$ is given by
\begin{equation}
 \label{eq_teff}
  T_{\rm eff} (r)
  = 6.2 \times 10^5 M_7^{-1/4}
   \brfrac{\Mdot}{\Ledd / c^2}^{1/4} 
   \brfrac{r}{\rs}^{-3/4}  {\rm K}.
\end{equation}
The simplest procedure to estimate $\lsg$ is to use the 
Wien displacement law assuming that the emission is close to a blackbody 
radiation: 
namely $ \lsg \simeq h c / 4 k \Teff$ 
(see Rybicki \& Lightman 1979).
It follows that the critical wavelength is
\begin{equation}
 \label{eq_lsg}
 \lsg = \frac{h c}{4 k \Teff(\rsg)} 
      = 0.36 \, \left[\frac{\Teff(\rsg)}{\rm K}^{-1}\right] {\rm cm}.
\end{equation}
The results are shown by 
the dotted lines in Fig.~\ref{fig:rlambda3} (lower panels).

The effects of electron scattering (opacity and Comptonization) 
must be considered in studying the emergent
spectra from a super-Eddington accreting disc 
(Shimura \& Manmoto 2003; Wang \& Netzer 2003; Kawaguchi 2003).
Comptonization is important at radius less than $\sim 300 \rs$ 
for $\Mdot \lesssim 1000 \Leddc2$,
and thus it has no effect 
on the emission beyond $\rsg$.
On the other hand, the ratio of electron-scattering opacity to 
absorption opacity is larger than unity at 
$r \lesssim 2500 [\Mdot/(\Leddc2)]^{2/3}$ (see Kato \ea 1998),
and thus the effect of electron-scattering opacity 
is still important at $\rsg$.

Due to the electron-scattering opacity, the emergent spectrum ($I_\nu$)
is distorted towards higher energy (i.e. shorter wavelength), 
and is close to the so called modified blackbody 
(Rybicki \& Lightman 1979;
Czerny \& Elvis 1987;
Wandel \& Petrosian 1988): 
\begin{eqnarray}
\label{eq_mbb}
 I_\nu = \frac{ 2 \, B_\nu(\Teff)}
  { 1 + \sqrt{ (\kappa_{\rm abs, \nu}+\kappa_{\rm es}) /
                     \kappa_{\rm abs, \nu} }}.
\end{eqnarray}
where $B_\nu$, $\kappa_{\rm abs, \nu}$ 
and $\kappa_{\rm es} (= 0.4\,$cm$^2$/g) are, respectively,
 the Planck function, absorption opacity and 
electron-scattering opacity. 
The spectral shift due to this effect is roughly estimated as
follows (Madej 1974):
\begin{eqnarray}
\label{eq_madej}
\Delta\log_{10}\lsg
\simeq \frac{1}{8} 
\log_{10}\left[\frac{\kappa_{\rm abs}(\rsg)}
          {\kappa_{\rm abs}(\rsg)+\kappa_{\rm es}(\rsg)}\right],
\end{eqnarray}
where $\kappa_{\rm abs}$ is the Rosseland mean opacity for
absorption (bound-bound, bound-free and free-free processes).
In assessing $\kappa_{\rm abs}$, we take 30 times the free-free
absorption opacity, 
as is done in Czerny \& Elvis (1987) and Kawaguchi (2003): 
\begin{equation}
\kappa_{\rm abs} = 30 \kappa_{\rm ff}
\end{equation}
with the free-free absorption opacity, $\kappa_{\rm ff}$, as follows
\begin{equation}
\kappa_{\rm ff} = 6.4 \times 10^{22} \, 
\left(\frac{\rho_{\rm mid-plane}}{{\rm g \, cm^{-3}}} \right) \, 
\left( \frac{T_{\rm mid-plane}}{{\rm K}} \right)^{-3.5} \, {\rm cm}^2/{\rm g}.
\end{equation}
Here, $T_{\rm mid-plane}$ is the temperature at the mid-plane of the disc 
(Kato \ea 1998).

Finally, $\lsg$ modified from Eq.~(\ref{eq_lsg}) by Eq.~(\ref{eq_madej})
are plotted in the lower panels of figure~\ref{fig:rlambda3} as 
solid lines.
As far as $\Mdot < \Mdot_{\rm ba}$, $\lsg$ is roughly 
proportional to $\Mdot^{-0.5}$.
For $\Mdot > \Mdot_{\rm ba}$ and  $\alpha=0.1$,
$\lsg$ reaches and stays at $\sim 4000 \rmAA$,
and interestingly it can be as small as $\sim 1500 \rmAA$ for
the lowest $\alpha$ case here (i.e. $\alpha =0.001$). 
Because of dust sublimation at $T \gtrsim 1500$ K, 
emission of dust falls within the Mid-IR--Near-IR spectral range 
and can not contribute to the optical component (see \S 3.3).
Therefore, emission from $\sim 2 \mu$m to $\lsg$ arises
from the self-gravitating part of the super-Eddington accreting disc.
In other words, this is 
{\it a discovery of a spectral window for observing the self-gravitating 
disc}, which has solely been studied by maser spots.

We note that $\lsg$ for $\Mbh \gtrsim 10^9 \Msun$ is also 
small even with sub-Eddington accretion rates.
In the sub-Eddington regime, however, 
heating by central radiation onto an outer
region at $r \sim \rsg$ may not be negligible.
An uncertainty of physical quantities (such as the gas density)
{ in the} outer region of the disc 
{ (e.g., H\=oshi \& Inoue 1988)} will lead to 
uncertainties of $\rsg$ and $\lsg$.
{ On the} contrary, a super-Eddington accreting disc is not flared, 
i.e. it has a maximum of 
the aspect ratio at $r \sim (5-100) \, \rs$ (Kawaguchi 2003), 
{ when such irradiation} is considered to be less important.
For this reason, $\rsg$ and $\lsg$ obtained for high $\Mdot$ 
are more reliable than for the sub-Eddington cases.
Therefore, we concentrate on the 
super-Eddington cases.

\section{Spectral Calculations}

We now demonstrate that the broad-band spectrum 
of a bright, well-studied NLS1, Ton~S~180, 
is indeed well fitted by the spectrum arising from 
three components: a slim disc (with a coronal, hard X-ray spectrum), 
a self-gravitating non-Keplerian disc 
and a dusty torus. 
The configuration we consider is schematically shown in Fig.~\ref{fig:config}.

The Near-IR to X-ray data are taken from Fig.~7 and Table~6 of
Turner et al. (2002).
The IRAS data for Far-IR and Mid-IR flux are derived from the NED database.
The luminosity is calculated from observed flux assuming 
 isotropic radiation, zero cosmological constant, 
 deceleration parameter $q_0 = 0.5$, and Hubble 
constant $H_0 = 75$  km/s/Mpc.
Data are plotted in Fig.~\ref{fig:fine_tune4}
as open squares. 
{ Ton S 180 has low Galactic and intrinsic extinction 
(Turner et al. 2002). 
Since Ton S 180 is a nearby object (z=0.06), 
we have not taken any spectral shifts or K-corrections into account.}

The BH mass estimated from the H$\beta$ width and optical 
luminosity (Wandel, Peterson \& Malkan 1999; Kaspi \ea 2000)
is $10^{7.1} \Msun$. This value is comparable to $10^{7.3} \Msun$ 
inferred from the [O III] width using Eq.~(2) in Wang \& Lu (2001).
Its B-band luminosity, $\nu \Lnu$(B), is about $10^{44.6}$erg/s,
implying  that $\Mdot \simeq 500 \Leddc2$ (see { Kawaguchi} 2003). 
Thus, this object is expected to be one of the 
highest $\Mdot / (\Leddc2)$ objects among NLS1s and Narrow-Line QSOs. 

\subsection{Spectrum of the inner, slim disc}

The structure and emergent spectrum of the inner disc around 
a non-rotating BH is computed 
numerically following the same procedure as in Kawaguchi (2003). 
The theoretical background of the numerical method is described in 
Matsumoto \ea (1984) and in Honma \ea (1991a,b). 
The model employs the standard prescription for turbulent viscosity 
$\nu_{\rm t}=\alpha c_{\rm s} H$ (Shakura \& Sunyaev 1973) where 
$c_{\rm s}$ is the sound speed and $H$ the semi-thickness.
The input parameters are $\Mbh$, $\Mdot$ and viscosity parameter $\alpha$.
Given a set of these parameters, we integrate the differential equations 
for the flow from $2 \times 10^4 \rs$ down to the inner free 
boundary at $1.01 \rs$.

\begin{figure*}
   \centering
    \includegraphics[angle=0,width=13cm]{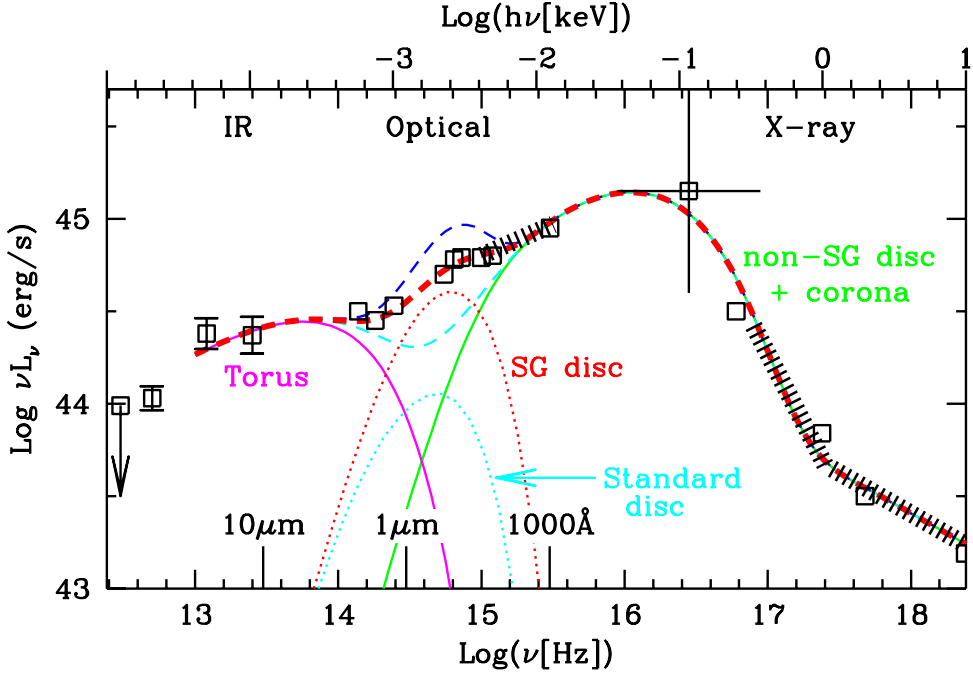}
   \caption{Mid-IR to X-ray spectral modeling of Ton~S~180.
    Open squares and thick striped lines are the observed data and 
    the power-law fits described in Turner \ea (2002).
    Two solid lines are the spectral components of the dusty torus 
    (left) and the non self-gravitating disc with a corona (right;
    $\Mbh = 10^{6.8} \Msun$, 
    $\Mdot = 1000 \Leddc2$, and $\alpha = 0.002$).
    Two dotted lines mean the spectrum (from $\rsg < r < r_{\rm out}$) 
    of the self-gravitating disc
    for $\gamma = -0.6$ (where $\Sigma \propto r^\gamma$) 
    { and $\alpha_{\rm out} = 0.02$} (upper; one of the successful
    fits shown in Table~\ref{tab:sg_fit}) and 
    { that of a Keplerian, standard disc} (lower).
    Total spectra of those three components are indicated 
    by dashed lines:
    { thick line is the spectrum with the self-gravitating 
    disc mentioned above, while the case with a standard disc is 
    indicated by a lower curve.
    The upper dashed line exhibits a total
    spectrum with a different parameter set for the outer 
    self-gravitating disc ($\gamma = 0, \ \alpha_{\rm out} = 0.02$).}
           }
              \label{fig:fine_tune4}
    \end{figure*}

The best parameters for this object have been selected by 
the least-square procedure, 
comparing the predicted spectrum 
(with a hard X-ray power-law due to a corona) to 
{ the power-law fits for the optical/UV/X-ray spectrum
(thick striped lines in Fig.~\ref{fig:fine_tune4}; 
see Table 7 in Turner \ea 2002)}.


First, 
{ the} following parameter region is surveyed;
$\Mbh = 10^{6.5-7.5} \Msun$, 
$\Mdot = 100-1000 \Leddc2$, and 
$\alpha = 0.1-0.001$ with 
$\Delta \log(\Mbh) = \Delta \log(\Mdot) = 0.5$,
and $\Delta \log(\alpha) = 1$.
Among them, one parameter set ($\Mbh = 10^7 \Msun$,
$\Mdot = 10^{2.5} \Leddc2$ and $\alpha = 0.001$) exhibits
the best fit.
Next, we compute the spectra in more detail around
this parameter set,
with $\Delta \log(\Mbh) = 0.2$,
$\Delta \log(\Mdot) = 0.25$,
and $\Delta \log(\alpha) = 0.3$.
Finally, we selected two best sets of parameters
{ as listed in Table~\ref{tab:best_nsg}, }
with $\alpha = 0.002$ in both cases.
{ 
The mass of this inner disc 
$M_{\rm disc}(r<\rsg)$ 
is 
$0.009 \times \Mbh$
for set 1, 
and $0.016 \times \Mbh$
for set 2, respectively.
Values at $\rsg$ ($\Sigma$ and $H$) are used 
as the inner boundary conditions for the 
self-gravitating part.
}

%
   \begin{table}
      \caption[]{Best parameter sets for the non self-gravitating disc.}
         \label{tab:best_nsg}
     $$ 
         \begin{array}{ccccccc}
            \hline
            \noalign{\smallskip}
	    {\rm set} & \Mbh [\Msun] & \rs [{\rm cm}] & \Mdot [\Leddc2]
	    & \rsg [\rs] & \Sigma [{\rm g / cm^2}]^{\mathrm{a}} 
	    & H [{\rm cm}]^{\mathrm{b}} \\
            \noalign{\smallskip}
            \hline
            \noalign{\smallskip}
	    1 & 10^7     & 3.0 \, 10^{12} & 10^{2.75} & 2000 
	    & 2.4 \, 10^6 & 1.2 \, 10^{15} \\
	    2 & 10^{6.8} & 1.9 \, 10^{12} & 1000              & 3000 
	    & 2.6 \, 10^6 & 1.3 \, 10^{15} \\
            \noalign{\smallskip}
            \hline
         \end{array}
     $$ 
\begin{list}{}{}
\item[$^{\mathrm{a}}$] Half surface density $\Sigma$ at $r = \rsg$.
\item[$^{\mathrm{b}}$] Semi thickness $H$ at $r = \rsg$.
\end{list}
   \end{table}
%


{ The emergent spectrum from $r < \rsg$ (for set~2) is shown
in Fig.~\ref{fig:fine_tune4} by a solid line (right),
while} 
relevant physical quantities are shown in Figure~\ref{fig:sotmdq}.
For a presentation purpose, a simple $\Teff$ profile 
(i.e. $\Teff \propto r^{-3/4}$) is also drawn as a { dotted} line.

{ As to the broken power-law description for the X-ray data,
different combinations of the photon index $\Gamma$ and breaking energy 
are deduced for different observations.
In order to assess how the different combinations
affect the generation of the best sets of parameters,
we have also performed the same least-square procedure with
the X-ray data replaced by the broken power-law description
in Vaughan \ea (2002).
The following four parameter sets provide best fits;
$[\log (\Mbh / \Msun), \ \Mdot/(\Leddc2), \ \alpha] = 
(6.8, 1000, 0.002), (6.8, 1000, 0.004), (7.0, 562, 0.002)$,
and $(7.2, 178, 0.002)$.
The goodness of the fit to the data 
shows a slightly shallower distribution than the fit to the 
Turner \ea (2002) data.
In the rest of this study, we emphasize the fit to 
the data in Turner \ea (2002), where the broadband data 
were obtained quasi simultaneously.
}

\subsection{Spectrum of the non-Keplerian, self-gravitating disc}

In the standard theory of radiatively cooled accretion discs 
(e.g., Pringle 1981), the effective temperature $\Teff$ 
(close to the surface temperature) is given by
\begin{equation}
\sigma \Teff^4(r) = \nu_{\rm t} \Sigma 
 \left(r \frac{d \Omega}{d r}\right)^2,
\label{eq:ts}
\end{equation}
where $\Omega$ is the rotation frequency. 
Again, the turbulent viscosity is chosen to be the standard 
prescription: 
$\nu_{\rm t}=\alpha c_{\rm s} H$ (Shakura \& Sunyaev 1973).
The sound speed $ c_{\rm s}$ is found from the requirement of hydrostatic
equilibrium, namely
\begin{equation}
c_{\rm s}^2 = \Omega^2_{\rm K} H^2 + 4 \pi G \Sigma H.
\label{eq:cs}
\end{equation}

At $r \gtrsim \rsg$, the disc becomes self-gravitating, 
namely $\Omega$ is expected to depart from its Keplerian 
value $\Omega_{\rm K}$. 
In order to compute $\Omega$, we assume that the rotation 
curve has only a gravitational origin, that is
\begin{equation}
r \Omega^2(r)=\frac{GM_{\rm BH}}{r^2}-g_R^{\rm disc}(r),
\label{eq:omega}
\end{equation}  
where $g_R^{\rm disc}$ is the contribution due to
the whole disc, including the inner slim disc.
In general, at large distances, we expect $g_R^{\rm
disc} < 0$ with the consequence that $\Omega > \Omega_{\rm K}$.
{ This means a possible reduction of the gradient,
$| d \Omega / d r|$, and subsequently a change of the effective
temperature [eq.~(\ref{eq:ts})] with respect to a standard disc.}
It is worth noting that a similar idea has been proposed by Lodato
\& Bertin (2001) in order to explain the IR-excess of T-Tauri stars. 

{ 
Since the self-gravitating disc should be gravitationally unstable, 
other sources of heating, e.g., heating by gravitational 
instabilities (Adams, Ruden \& Shu 1988; Lodato \& Bertin 2001),
may overcome the heating by turbulent viscosity (see Lodato \& Rice 2003). 
To take into account this possibility, we allow the parameter $\alpha$
 in the self-gravitating part to increase linearly with $r$,
from $0.002$ (the value used for the inner disc in \S~3.1)
at $\rsg$ to $\alpha_{\rm out}$ at the outermost radius of the
self-gravitating disc $r_{\rm out}$.
} 

Since there is no simple formula for radial self-gravity, we
 have determined $g_R^{\rm disc}$ numerically using the accurate Poisson
3D-solver, which is described in Hur\'e (2003) for discs with various
shapes, sizes and surface density profiles. 
This has been performed assuming
 that $H \propto R^\beta$ and $\Sigma \propto R^\gamma$, 
where $\beta$ and $\gamma$ are input parameters.
Here, $H$ and $\Sigma$ are continuous at the transition between 
the two regimes of the disc.
{ The radial extension of the self-gravitating disc 
$r_{\rm out}$ and the viscosity parameter $\alpha$ at that radius
$\alpha_{\rm out}$ are also input parameters.
Those values ($\beta$, $\gamma$, $r_{\rm out}$, and $\alpha_{\rm out}$)
are determined so that the total spectra (\S 3.4) provide acceptable
fits to the observed one.}

For each model, the effective temperature $\Teff(r)$ is 
derived from the resultant rotation frequency using Eq.~(\ref{eq:ts}).
The spectrum of the self-gravitating disc is then computed 
{ in the same way as the inner slim disc (i.e., with
the effect of electron scattering, even though the effect is small
in the outer self-gravitating part).}

{ 
When we add the spectra of the
outer, self-gravitating disc onto the inner disc spectrum,
it is found that 
the combined spectra based on set~2 (Table~\ref{tab:best_nsg}) 
generally provide a better fit
to the observed optical/UV spectrum than those based on 
set~1.
Thus, 
we hereafter discuss the models obtained for set~2 alone 
as an example. 
}

{ For set~2, three possible combinations of parameters
are shown in Table~\ref{tab:sg_fit}.
The best parameters for $\beta$ and $r_{\rm out}$ are common to
them: $\beta \simeq 1$ and $r_{\rm out} \simeq 3 \times 10^4 \rs \ 
(= 10 \times \rsg)$.
For one case ($\gamma = -0.6$ and $\alpha_{\rm out} = 0.02$),
the spectra from the self-gravitating part (upper dotted line)
and the total spectrum (thick dashed line) are presented in 
Fig.~\ref{fig:fine_tune4}.
Those three combinations exhibit almost identical spectra 
(deviation of $\nu L_\nu$ is $0.05$ dex at most).
Other combinations are also acceptable.
The cumulative disc mass
$M_{\rm disc}(r \le r_{\rm out})$ is not negligible, 
ranging from $0.4 \Mbh$ to $7.0 \Mbh$. 
For the intermediate case, for instance, 
the total mass, $\Mbh + M_{\rm disc}$, 
is about $2.4 \, \Mbh = 10^{7.2} \Msun$, which is consistent
with the central mass estimated by H$\beta$ and [O III]
line widths.
It indicates that $\Mbh$-estimations using line widths 
overestimate the BH mass systematically.}
To assess how the spectrum changes with different $\gamma$, 
the total spectrum obtained with $\gamma = 0$, which is less preferable, 
is indicated { by the upper dashed line}.

%
   \begin{table}
      \caption[]{Successful solutions for the self-gravitating disc.}
         \label{tab:sg_fit}
     $$ 
         \begin{array}{rll}
            \hline
            \noalign{\smallskip}
          \gamma \hspace*{1mm}& \hspace*{6mm}\alpha_{\rm out} 
	    & M_{\rm disc}(r \le r_{\rm out}) / \Mbh \\
            \noalign{\smallskip}
            \hline
            \noalign{\smallskip}
           0.3 & \hspace*{5mm}0.002^{\mathrm{a}} & \hspace*{7mm}7.0     \\
	  -0.6 & \hspace*{5mm}0.02               & \hspace*{7mm}1.4     \\
          -1.5 & \hspace*{5mm}0.1                & \hspace*{7mm}0.4     \\
            \noalign{\smallskip}
            \hline
         \end{array}
     $$ 
\begin{list}{}{}
\item[$^{\mathrm{a}}$] Namely, constant $\alpha$ in the entire disc.
\end{list}
   \end{table}

\begin{figure}
   \centering
    \includegraphics[width=8.5cm]{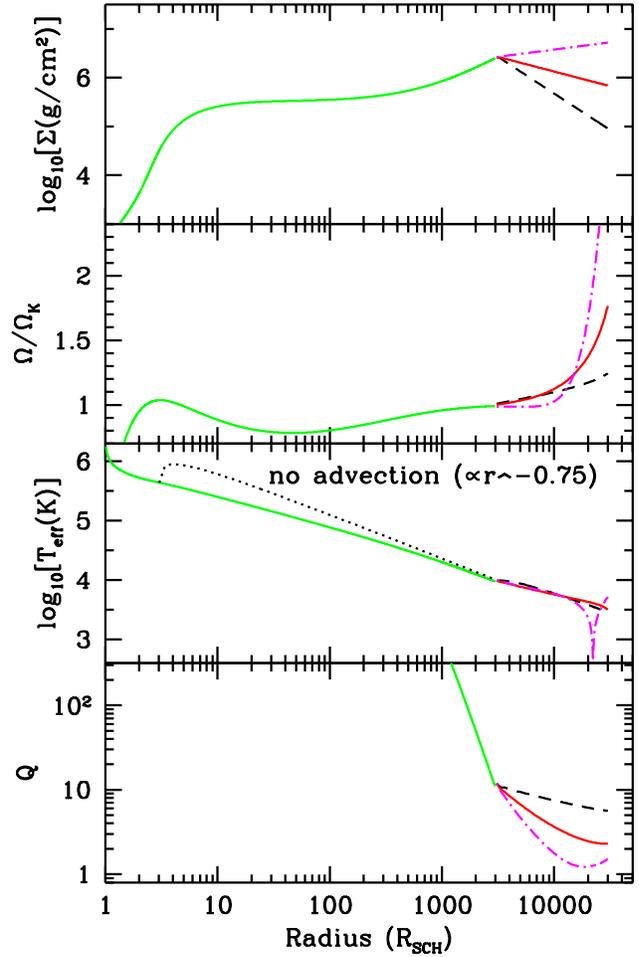}
    \caption{Variation with radius of some disc parameters 
    { (the inner, non self-gravitating part is based on set~2)}; 
    half surface density $\Sigma$, rotational frequency 
    normalized to the Keplerian value $\Omega / \Omega_{\rm K}$,
    effective temperature $\Teff$, 
    and Toomre $Q$-parameter.
    The solid lines are computed with $\gamma = -0.6$ and 
    $\alpha_{\rm out} = 0.02$, while dot-dashed and dashed ones 
    are computed for $\gamma=0.3$ with a constant $\alpha$ and
    $\gamma = -1.5$ with $\alpha_{\rm out} = 0.1$, respectively.
    Two accretion regimes are detached at $r = \rsg (= 3000 \rs)$.
    }
         \label{fig:sotmdq}
   \end{figure}

Figure~\ref{fig:sotmdq} presents several physical quantities 
of the disc 
{ for the three successful parameter sets (Table~\ref{tab:sg_fit})}
as a function of radius; 
$\Sigma$, $\Omega$ (Eq.~\ref{eq:omega}) and $\Teff$.
{ Vertical optical thickness of the flow, $\tau$ 
[$\tau \, = \, \Sigma \times (\kappa_{\rm abs} + \kappa_{\rm es}) 
 \, > \, \Sigma \, \kappa_{\rm es} \, = \, 
 0.4 \, (\Sigma / {\rm g \, cm^{-2}})$] is much larger than unity
everywhere.
For the case with a very massive disc ($\gamma = 0.3$), 
$\Omega$ becomes flat and even starts to increase with radius
due to the strong radial self-gravity.
This produces a local decrease in $\Teff$ [$\propto (d \Omega / d r)^{0.5}$;
eq.~(\ref{eq:ts})] 
at a small range of radius.
However, the temperature profiles $\Teff (r)$ for the three
cases are mostly the same, and thus they produce almost identical
spectra.}

%
%

\subsection{Emission from the dusty torus}

The dusty torus component
is supposed to be responsible for the Mid-IR ($\sim 30 \mu$m) 
to $\sim 2 \mu$m  emission.
In general, the Near-IR flux from AGN varies with time following the 
optical flux with a time lag of about a month--year (e.g., 
Clavel, Wamsteker \& Glass 1989; Glass 1992; Nelson 1996).
This implies that the dusty torus is located quite far from the BH,
and that it is heated by the radiation from
the vicinity of the central BH.

A detailed modeling of the emission of the torus is beyond
the scope of this paper. Here, we adopt a simple power-law 
description with a cut-off. Theoretical arguments on the dust 
sublimation as well as 
observational data (as a spectral bump around 3 $\mu$m) 
indicate that the maximum temperature of the torus 
 is $\sim 1000-1500$K (e.g. Efstathiou, Hough \& Young 1995; 
Kobayashi \ea 1993; Pier \& Krolik 1993).
It corresponds to wavelength of $\sim 2.4 \mu$m 
[i.e. $h c / (4 k T)$ with $T = 1500 {\rm K}$], 
or $\nu \simeq 10^{14.1}$ Hz (i.e. $4 k T / h$).

We assume that the fluxes at $25 \mu m$ and $12 \mu m$ come from 
the dusty torus alone, while the near-IR flux is partly due to 
the self-gravitating disc. 
Then, the simple model we apply is
 \begin{equation}
\nu \Lnu = 10^{38.45} \nu^{0.45} \exp(- \nu / 10^{14.1} {\rm Hz}) 
 \, {\rm erg/s}.
\end{equation}
We are confident that this is neither the only nor the best 
possible description. 
The spectral index inferred for this object differs
from the typical mid-IR--near-IR index of bright AGNs 
($\nu \Lnu \propto \nu^{-0.4}$; Neugebauer \ea 1987; 
Polletta \& Couvoisier 1999; see also Grupe \ea 1998).

We can specify the location $r_{\rm t}$ of the inner torus 
(see Fig~\ref{fig:config}) by considering that the torus is 
in thermal balance (heating by radiation from the inner disc 
is balanced by radiative cooling):
\begin{equation}
r_{\rm t} \simeq 5.3 \times 10^{17} \, 
\brfrac{L}{10^{45} \, {\rm erg/s}}^{0.5}
\brfrac{T}{1500 \, {\rm K}}^{-2} \, {\rm cm}.
\end{equation}
where $L$ and $T$ are the bolometric luminosity of the central
disc, and the gas temperature at the innermost radius of the torus,
respectively.
For the set~2, we have $r_{\rm t} \simeq 5.3 \times 10^{17} {\rm cm} \,
(=0.17\,{\rm pc}) \simeq 2.8 \times 10^5 \rs$, which is about 10 times 
larger than the outer radius of the self-gravitating disc $r_{\rm out}$.
Thus, gravitational attraction on the self-gravitating disc by 
the torus should not be crucial, unless the torus mass exceeds
the BH mass by some order of magnitude.

\subsection{Mid-IR -- X-ray spectral energy distribution}

{ The thick dashed line in}
Figure~\ref{fig:fine_tune4} shows the broad-band, mid-IR to X-ray
spectral modeling for Ton~S~180,
{ comprised of the three components described in the
subsections above}. 
We see that the
 optical component can be interpreted as the thermal emission 
from the self-gravitating, non-Keplerian disc.
{ The models for the self-gravitating disc with 
$\gamma$ and $\alpha_{\rm out}$ listed in Table~\ref{tab:sg_fit}
provide equally good fits to the observed spectrum.
The upper dashed line (for $\gamma = 0$ and $\alpha_{\rm out} = 0.02$) 
is drawn to show to what extent a different
value of $\gamma$ changes the total spectrum.
}

If the outer self-gravitating disc is replaced by 
a Keplerian, standard disc, the resultant total 
spectrum (lower dashed curve)
 is not adequate to explain the observed optical/near-IR spectrum.
{ In other words, flux enhancement at the outer region
is necessary.
This enhancement depends on the resultant rotation law,
as well as $\alpha (r)$, 
at the outer self-gravitating disc.
}

{ 
We demonstrate that a broadband spectrum of one NLS1, Ton~S~180, 
is indeed well fitted by a summation of 
an inner, non self-gravitating disc, an outer, self-gravitating
disc and a dusty torus.
Detailed mid--near-IR observations, including temporal
studies (e.g. Glass 1992; Nelson 1996), will 
provide a more accurate model of the dusty torus.
Subtraction of the torus component from the total (disc and torus) 
spectrum will enable us to 
derive stronger constraints on the size of the self-gravitating
part, and mass of the whole disc, etc.
High spatial resolution obtained with ground-based, 
mid- to near-IR interferometers with the large 
telescopes, such as OHANA (Mariotti \ea 1996; Perrin \ea 2000)
and VLTI, will contribute to solving these problems.
}

\section{ Implications and limitations of the present study}


{ As is obtained in \S 3.2, }
the integrated disc mass is about $(0.4-7.0) \Mbh$ (Tab.~\ref{tab:sg_fit}).
Although 
the Toomre $Q$ parameter
$[=\Omega c_s/ (\pi \Sigma G)]$ remains greater than one 
in the whole disc (Fig~\ref{fig:sotmdq}),
{ it is quite close to unity at outer radius for larger $\gamma$.
This fact supports the introduction of the additional heating
due to gravitational instabilities (e.g., increasing $\alpha$ used
here; see Lodato \& Rice 2003).}
{ Consensus on the amount of such heating has not yet been reached.
Further understanding on this topic, which is related to $\alpha_{\rm out}$
in this study, will allow us to constrain $\gamma$ 
(where $\Sigma \propto r^\gamma$) and the disc mass more severely.
}

The size of the broad-line region (BLR), $R_{\rm BLR}$, for Ton~S~180 is 
expected to be of the order of 100\,lt.d.,
using the empirical relation between the optical luminosity 
and $R_{\rm BLR}$ (Kaspi \ea 2000).
This radius corresponds to 0.085\,pc $= 2.6 \times 10^{17}$\,cm; 
equivalently, $1.4 \times 10^5 \rs$ for $\Mbh = 10^{6.8} \Msun$.
Thus, the broad-line region is likely located 
between the inner radius of the torus 
($r_{\rm t} \simeq 3 \times 10^5 \rs$)
and the outer radius of the self-gravitating disc 
($r_{\rm out} \simeq 3 \times 10^4 \rs$).
However, the constraint on $r_{\rm out}$ 
derived in \S~3.2 is rather weak.
If the outer massive disc extends at $r \gg r_{\rm out}$,
the clouds in the broad-line region may not be in 
Keplerian motion, which will lead to an uncertainty
for the $\Mbh$-estimation with H$\beta$ width (e.g., Krolik 2001).

Since the main purpose of this paper is to report 
the discovery of the spectral window for observing a
self-gravitating disc, 
and to demonstrate a spectral fit to the observed spectrum 
of a super-Eddington accreting AGN,
we did not treat the following two issues very accurately:
(i) the vertical self-gravity (the last term in the 
right-hand-side of Eq.~\ref{eq:cs}) in the inner, non self-gravitating disc,
and (ii) the effect of realistic absorption opacity in deriving
$\rho_{\rm mid-plane}$ (\S 2.1).
{ Also, possible outflow from the disc is not taken into
account either in the inner slim part or in the outer self-gravitating
part.
Outflow/evaporation at the outer region may contribute to the 
formation of BLR clouds or broad absorption line (BAL) clouds.}
These issues will be examined in the future.

%

\section{Conclusions}

We have presented the 
outermost radius of the non self-gravitating accretion disc ($\rsg$)
around a super-massive black hole (BH), i.e.
the inner edge of the self-gravitating disc,
as a function of the BH mass $\Mbh$, accretion rate $\Mdot$ and 
viscosity parameter $\alpha$. 
We then showed the 
corresponding wavelength ($\lsg$) that is relevant to the 
emission from $\rsg$.

When 
$\Mdot \lesssim 70 \, \Leddc2$ (for $\Mbh = 10^7 \Msun$),
$\lsg$ is roughly 
proportional to $\Mdot^{-0.5}$ for fixed $\Mbh$ and $\alpha$.
With a higher $\Mdot$, 
$\lsg$ reaches and stays at $\sim 4000 \rmAA$ for $\alpha=0.1$.
Interestingly, $\lsg$ is as small as $\sim 1500 \rmAA$ for
the lowest $\alpha$ case in this study (i.e. $\alpha =0.001$). 
Therefore, the continuum emission from $\sim 2 \mu$m to $\lsg$ arises
from the self-gravitating part of the super-Eddington accreting disc.
Thus, we have discovered a spectral window for observing the 
self-gravitating disc, which has only been studied by maser spots
at the parsec-scale for a few nearby objects.

Next, we demonstrated that the mid-IR to X-ray spectrum 
of a bright, well-studied Narrow-Line Seyfert 1 galaxy, Ton~S~180, 
is indeed well fitted by the spectrum arising from 
the following three components: an inner slim disc (with a corona), 
an outer, self-gravitating non-Keplerian disc 
and a dusty torus. 
{ Comparing the model spectra of the slim disc with the 
observed UV--X-ray one, the following parameters are favored:
$\Mbh \approx 10^{6.8} \Msun$, $\Mdot \approx 1000 \Leddc2$,
and $\alpha \approx 0.002$.}
In the model for the outer, self-gravitating disc, 
{ we allow the viscosity parameter $\alpha$ to increase 
with radius, and }
an outer radius of $3 \times 10^4 \rs$
is inferred.
{ Various profiles for the surface density $\Sigma$ are
acceptable, e.g., from $\Sigma \propto r^{0.3}$ to 
$\Sigma \propto r^{-1.5}$, depending on $\alpha$ at the outermost radius
of the disc.}

{ The accretion disc is quite massive, and the disc mass
can be $(0.4 - 7.0) \Mbh$.}
The total mass, BH mass plus the entire disc mass, 
is found to be about $2.4 \, \Mbh = 10^{7.2} \Msun$
{ if $\Sigma \propto r^{-0.6}$}, which is consistent
with the central mass estimated by 
{ H$\beta$ and [O III] line widths. 
This indicates that $\Mbh$-estimations using line widths 
 systematically overestimate the BH mass.}
Although the disc mass is quite high relative to the BH mass, 
the Toomre $Q$ parameter remains greater than one in the whole disc, 
so that the disc is marginally stable. 

Moreover, 
$\rsg$ can be less than the photon trapping radius 
for $\Mdot \gtrsim 10^{3.3} \Leddc2$.
Thus, a self-gravitating, 
optically-thick, advection-dominated accretion disc
is expected to appear
in the extremely high accretion rate regime.

\begin{acknowledgements}

We thank 
Suzy Collin, 
Helene Sol, 
Julien Woillez, 
Catherine Boisson,
and Shin Mineshige for useful comments. 
{ Detailed and productive comments from the 
referee, Anuradha Koratkar, are also appreciated.}
We are deeply grateful to Ryoji Matsumoto and Fumio Honma 
who developed 
the numerical code to solve the radial structure of 
super-Eddington accretion discs.
{ This research has made use of the NASA/IPAC 
Extragalactic Database (NED) 
which is operated by the Jet Propulsion Laboratory,
California Institute of Technology, under contract with the National 
Aeronautics and Space Administration.} 
T.K.\ is supported by the Japan
Society for the Promotion of Science (JSPS) 
Postdoctoral Fellowships for Research Abroad (464).

\end{acknowledgements}

\end{document}